# Heat conduction tuning using the wave nature of phonons

Jeremie Maire[*,1,2], Roman Anufriev[1], Ryoto Yanagisawa[1], Aymeric Ramiere[1,2], Sebastian Volz[3] and Masahiro Nomura[*,1,4,5]

[1]Institute of Industrial Science, The University of Tokyo, Tokyo 153-8505, Japan

[2]Laboratory for Integrated Micro Mechatronic Systems/National Center for Scientific Research-Institute of Industrial Science (LIMMS/CNRS-IIS), The University of Tokyo, Tokyo 153-8505, Japan

[3]Laboratoire d'Energétique Moléculaire et Macroscopique, Combustion UPR CNRS 288, Ecole Centrale Paris, Grande Voie des Vignes F-92295 Chatenay-Malabry, France

[4]Institute for Nano Quantum Information Electronics, The University of Tokyo, Tokyo 153-8505, Japan

[5]PRESTO, Japan Science and Technology Agency, Saitama 332-0012, Japan

*e-mail: *nomura@iis.u-tokyo.ac.jp, jmaire@iis.u-tokyo.ac.jp*

**The world communicates to our senses of vision, hearing and touch in the language of waves, as the light, sound, and even heat essentially consist of microscopic vibrations of different media. The wave nature of light and sound has been extensively investigated over the past century[1,2] and is now widely used in modern technology. But the wave nature of heat has been the subject of mostly theoretical studies[3,4], as its experimental demonstration, let alone practical use, remains challenging due to the extremely short wavelengths of these waves. Here we show a possibility to use the wave nature of heat for thermal conductivity tuning via spatial short-range order in phononic crystal nanostructures. Our experimental and theoretical results suggest that interference of thermal phonons occurs in strictly periodic nanostructures and slows the propagation of heat. This finding broadens the methodology of heat transfer engineering by expanding its territory to the wave nature of heat.**

Interference – one of the most remarkable wave phenomenon – can be demonstrated in periodic structures, where systematic reflections of waves result in their constructive and destructive interference. Specifically designed periodic structures can even fully control the propagation of light (photons) or sound (phonons) and thus are called photonic[5,6] or phononic crystals[7,8]. But application of this concept to manipulation of heat (ensemble of thermal phonons) requires nanoscale periodicity and a structure with exceptionally smooth interfaces[3]. Indeed, wavelengths of thermal phonons at room temperature are only a few nanometres[9,10] and phonons quickly lose their phase (coherence) being scattered at surfaces with nanometre roughness[11]. For this reason, room temperature coherent scattering of thermal phonons (i.e. scattering with preserved phase) has been demonstrated only in superlattices with atomically-smooth interfaces and nanometre-size periodicity[12,13]. However, many of the practical applications of this phenomenon, such as waveguides or cloaking, cannot be realized in superlattices and require two-dimensional (2D) structures[4], similar to those used in photonics to manipulate light[5,6]. Such 2D phononic crystals can be fabricated via lithography techniques, but the atomically smooth surfaces and short periodicity are challenging to achieve.

In the past decade, advances in nanofabrication[14] resulted in a burst of experimental demonstrations of the reduction in thermal conductivity of 2D phononic crystal structures, which was attributed to the interference of thermal phonons[15–17]. But some theoretical works[10,18,19] recognized that this reduction originates from factors unrelated to phonon interference (e.g. surface roughness, strong surface scattering). Thus, the low absolute thermal conductivity alone cannot be an indication of phonon interference and comparative studies are required to detect the precise impact of such phenomenon.

Moreover, the periodicity is typically set for a specific and rather long wavelength, but, unlike light or sound, heat essentially consists of thermal phonons in a broad range of rather short wavelengths[10]. Nevertheless, recent theoretical studies on 2D phononic crystals have demonstrated that phonon interference can actually affect the entire phonon spectrum at low temperature, reducing overall phonon group velocity and density of states[20,21], whereas experiments at sub-kelvin temperatures have shown that, if phonon wavelengths are sufficiently long, thermal conductance can be controlled by phonon interference, despite the micrometre scale of the phononic crystals[20,22]. Still, the temperature range of impact of these interferences had not been clarified, as the possibility of such an effect at room temperature remains controversial[10,18].

Here we present a parallel study on nanoscale-sized 1D and 2D phononic crystals and demonstrate that thermal conductivity is reduced in phononic crystals with an ordered array of holes as compared to that with randomly positioned holes due to the presence of phonon interference, opposite to theoretical predictions[23]. Moreover, we show the temperature dependence of this effect and observe the transition from coherent to incoherent heat transport, thus answering the long-lasting question about the range of occurrence of such phonon interference. In addition, we perform finite element (FEM) and Monte-Carlo simulations to show that our observations cannot be attributed to incoherent scattering mechanisms.

We study phononic crystals with 1D (Fig. 1a) and 2D (Fig. 1b) arrays of holes. All samples were fabricated on a SOI wafer with a 145 nm-thick single-crystalline top silicon layer (Methods). Each sample consisted of a suspended $5 \times 5$ μm² silicon island topped by a $4 \times 4$ μm² aluminum pad and supported by five nano-beams ($0.3 \times 10$ μm²) or a membrane ($5 \times 15$ μm²) on each side. The phononic crystals, with a period of 300 nm, were formed in the nano-beams and membranes and consisted of arrays of holes with diameters of 161 and 133 nm in 1D and 2D phononic crystals, respectively. A set of samples consisted of the structures with degrees of disorder ($\delta$) in the 0 – 14% range (Fig. 1), where the shift of each hole from its aligned position is given by $\varepsilon \cdot \delta \cdot 300$ nm, with $\varepsilon$ as a random number in the range from −1 to +1. Since each set of samples was fabricated simultaneously on the same wafer, the surface roughness, which was estimated to be approximately ~2.5 nm (see details in Supplementary Information), did not vary from one sample to another.

To study the in-plane heat conduction in the phononic crystals, we used a micro-TDTR technique[24], originally designed for suspended micrometer-sized membranes (see the Supplementary Information). The schematic of the experimental setup is shown in Fig. 1c. The samples are mounted in a He-flow cryostat. A pulse laser beam (642 nm) periodically heats the aluminum pad in the centre of the sample, while a continuous laser beam (785 nm) constantly measures relative change of its reflectivity (ΔR/R) caused by the increase in temperature. As heat gradually dissipates after each heating pulse, the reflectivity returns to its initial value. The measured time (t) dependence of this heat dissipation (Fig. 1c) can always be well fitted by an exponential decay $\exp(-\gamma t)$, with γ as the thermal decay rate – the only parameter characterizing heat conduction in each sample. The standard deviation of the thermal decay rate is calculated from its fluctuations over time and remains less than 2%.

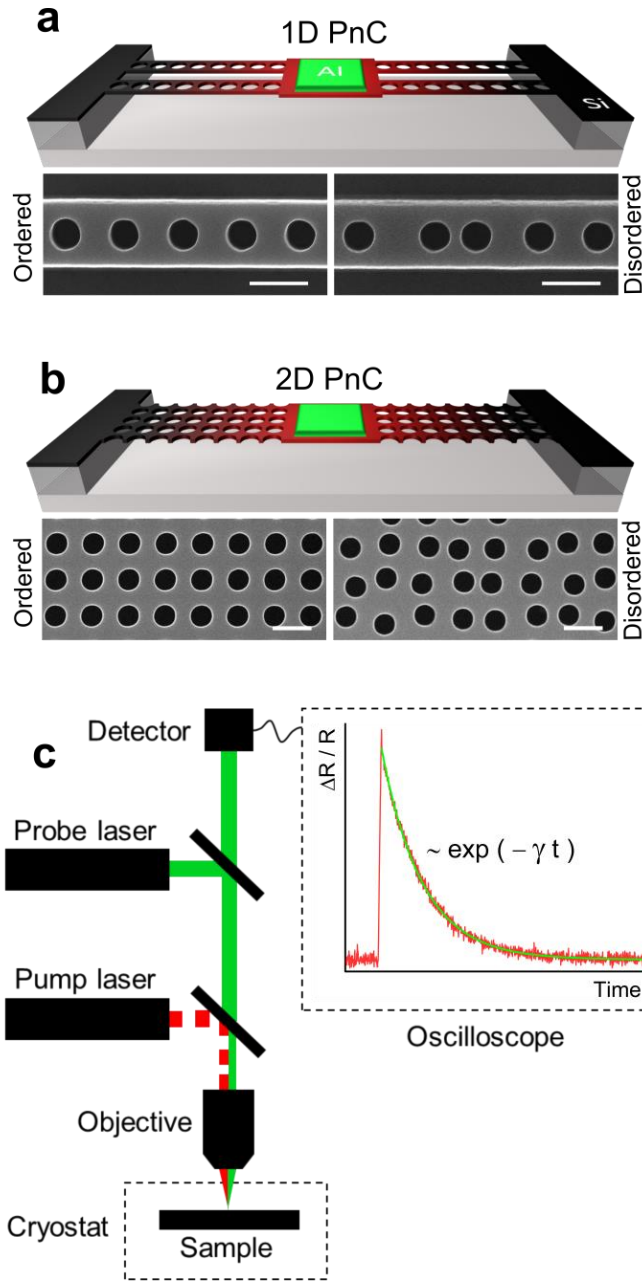

**Figure 1 | Samples and experimental setup**. Schematic and SEM images show fabricated samples of 1D (**a**) and 2D (**b**) phononic crystals with ordered ($\delta = 0\%$) and disordered ($\delta = 15\%$) arrays of holes. Scale bars are 300 nm. Schematic of the μ-TDTR setup (**c**) with inset showing a typical thermal decay curve with an exponential fitting.

First, we investigate the impact of the disorder on heat conduction in phononic crystals at the temperature of 4 K (Fig. 2). In the samples with significantly disordered positions of holes ($\delta > 5\%$) heat dissipates at the same rate, regardless of the disorder degree. But in the samples with small disorder degrees ($\delta < 5\%$) heat dissipation slows down, becoming 8.5% and 19% slower in perfectly ordered 1D and 2D phononic crystals, respectively. In terms of thermal conductivity, the reduction in ordered structures reaches 8.5% and 21%, respectively (see Supplementary Information), thus the short-range order of the phononic crystal can be used to tune the thermal conductivity of nanostructures.

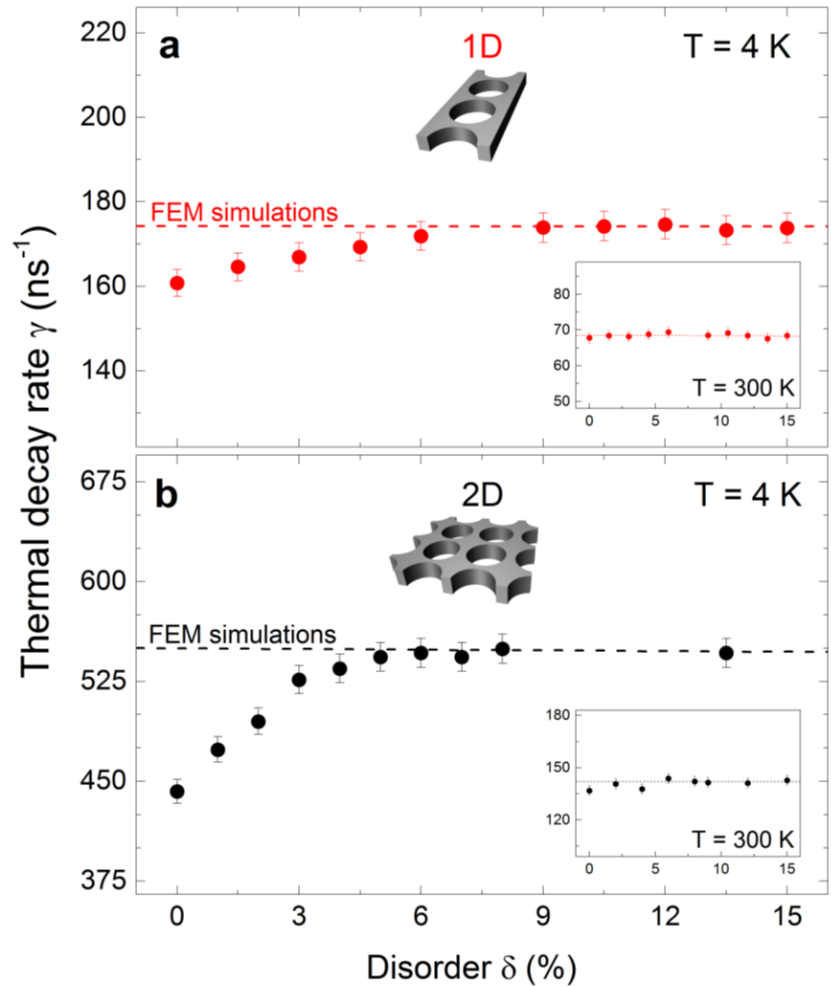

**Figure 2 | Thermal decay rate measurements with varying disorder.** Measured thermal decay rates in both 1D (**a**) and 2D (**b**) ordered phononic crystals deviate from that of disordered structures at 4 K, whereas at 300 K (insets) heat dissipates through ordered and disordered structure at an equal rate. Error bars show a standard deviation during the measurements (also included into the points at 300 K). Dashed lines show results of the FEM simulations based on the Fourier heat transport equation.

At room temperature, however, this reduction is absent and the decay rates seem to be independent of the disorder (insets of Fig. 2), with average thermal conductivities of 33 Wm$^{-1}$K$^{-1}$ for 1D and 42 Wm$^{-1}$K$^{-1}$ for 2D phononic crystals. We performed FEM simulations based on the Fourier heat transport equation, keeping all parameters, except the disorder degree, identical. The results, displayed as dashed lines in Fig. 2, show no impact of disorder in the Fourier heat transport limit (see details in Supplementary Information).

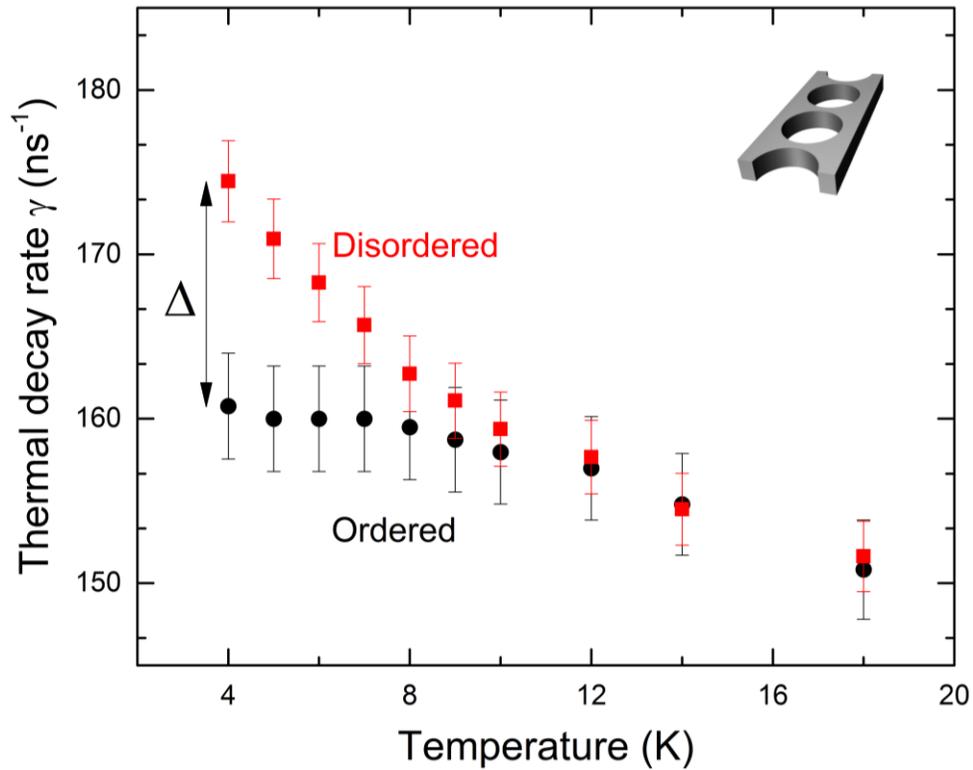

**Figure 3 | Temperature dependence of thermal decay rates in 1D PnC.** The decay rate in the disordered case ($\gamma_{disordered}$, calculated as the average between disorders of 10.5 and 12%) increases as temperature is decreased, whereas the decay rate in the periodic ($\gamma_{ordered}$) sample deviates from this trend. Error bars show a standard deviation during the measurements.

Next, we investigate the temperature dependence in more details, focusing on the 1D phononic crystals for simplicity. In Fig. 3, we compare the temperature dependences of decay rates in the phononic crystals with perfectly aligned ($\delta = 0\%$) and disordered arrays of holes. In the disordered system (average between structures with $\delta = 10.5\%$ and $\delta = 12\%$), cooling gradually increases the decay rate. But in the ordered system the decay rate deviates from this trend below 10 K. Regarding the difference between these decay rates, we see a gradual decrease from the value of 8.5%, observed at 4 K, towards the absence of difference, as was observed at 300 K. A similar trend is found for 2D phononic crystals (see Supplementary Information, Fig. S5). These experimental results imply that phonon interference occurs in the ordered phononic crystals at low temperatures, where the phonon wavelengths can reach tens of nanometres, enabling coherent reflections of thermal phonons. But, as the temperature is increased, the wavelengths shorten, with a trend inversely proportional to the temperature, and the effect disappears as the coherent part of the phonon spectrum becomes negligible even for the ordered structure.

To show that this hypothesis quantitatively explains our results, we have to assume that only a low-frequency part of the phonon spectrum is affected by phonon interference, and that this part depends on the disorder degree. Indeed, Wagner et al.[25] recently experimentally demonstrated that coherent phonon modes could exist up to 50 GHz in their ordered 2D phononic crystals, but only up to 20 GHz in the disordered one. The cut-off frequency, below which all modes are coherent, was empirically determined in phononic crystals with a period of 300 nm as $f_c$ (R) = 0.043 $v_L$/R, where $v_L$ = 8433 m/s is the frequency-independent longitudinal sound velocity in silicon[25,26], and R is the parameter taking into account both the average surface roughness ($\sigma = 2.5$ nm) and the average displacement of the holes from their ordered position. The average displacement is taken as half the maximum displacement of the holes in any direction (maximum displacement: $\delta \times 300$). We calculate R as = $(\sigma^2 + (0.5 \times \delta \times 300)^2)^{1/2}$. For example, for our ordered structure (i.e. R = 2.5 nm), this empirical equation yields a cut-off frequency of $f_c$ = 145 GHz.

Since coherent modifications of the phonon dispersion dramatically suppress thermal conductance[20,21], we can assume that phonons below the cut-off frequency nearly do not contribute to heat conduction. We thus assume that the predominant contribution to heat conduction is carried by phonons above the cut-off frequency. This part of the spectrum can be approximated by the Planck distribution for phonons (see Supplementary Information). As the disorder is introduced, the cut-off frequency shifts to lower frequencies and localized modes become propagating, increasing the thermal conductance.

Then, the experimentally measured difference between ordered and strongly disordered structures, calculated as $\Delta = (\gamma_{ordered} - \gamma_{disordered}) / \gamma_{disordered}$, is simply equal to the ratio of the phonon's contribution to heat flux above the cut-off frequency to the entire spectrum (see complete derivation in Supplementary Information). Fig. 4a shows that this model, using a single experimentally determined coefficient, quantitatively agrees with the experimental data for 1D phononic crystals: $\Delta$ is about 13% in the ordered structure and gradually decreases with disorder due to the reduction of the cut-off frequency, in agreement with the experimental results. Above $\delta = 6\%$, the effect becomes negligible as the proportion of affected phonons becomes too small to contribute significantly to heat transport.

Next, we compare the temperature dependence obtained experimentally to that expected from the theory. Fig. 4b shows that the theoretical curve is again in a good agreement with our experimental data. Indeed, at temperatures above 4 K, the phonon population shifts to frequencies of hundreds of gigahertz and the reduction in thermal conductance due to the coherent phonon modes below 145 GHz, no longer significantly impacts the total heat conduction. For this reason, we observed the weakening of the effect as temperature increases, and finally detected nearly no impact of the disorder on the thermal conductivity above 10 K, in agreement with estimations by Marconnet et al.[27]. The absence of impact of disorder stays all the more valid at room temperature, in agreement with Wagner et al. [25].

To show that incoherent scattering mechanisms cannot explain the experimental results, we also performed 2D Monte-Carlo simulations of phonon transport in our 1D phononic crystals with different degrees of disorder (see Supplementary Information for details and 2D data). These simulations take into account impurity, boundary and phonon-phonon scattering, but not phonon interference. The simulations, using the same geometry as our experimental samples, show no impact of disorder at all temperatures (Fig. 4a). Thus, only coherent phonon scattering remains to explain the experimental results.

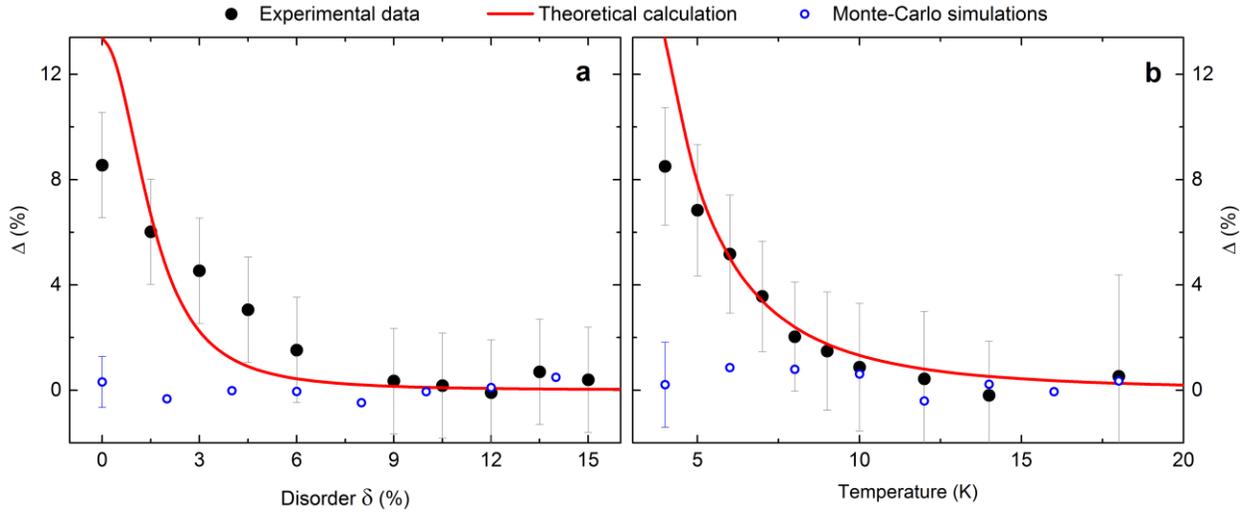

**Figure 4 | Comparison between experiment and theories. a** Theoretically expected disorder dependence alongside the experimentally measured difference between thermal decay rates ($\Delta = (\gamma_\delta - \gamma_{disordered}) / \gamma_{disordered}$). Disorder dependence of the difference $\Delta$ in the Monte-Carlo simulations (blue scatters). **b** Temperature dependence of the same effect, calculated as ($\Delta = (\gamma_{ordered} - \gamma_{disordered}) / \gamma_{disordered}$), for both experiments and theory. Temperature dependence of the difference $\Delta$ in the Monte-Carlo simulations (Blue scatters). Error bars show an upper bound of the signal deviation during the measurement.

In conclusion, our experimental data clearly showed that heat conduction can be tuned using the wave nature of phonons; it is hindered in perfectly periodic phononic crystals, in contrast with disordered structures. Our simulations showed that this effect cannot be attributed to incoherent scattering mechanisms, but assuming coherent scattering of the low-frequency phonons, we obtained a good agreement between the experiment and the theory. Furthermore, whereas Zen et al.[20,22] showed coherent reduction of thermal conductance at sub-kelvin temperatures, here we demonstrated thermal conduction control based on this coherence at one order higher temperatures, until the transition to purely diffusive heat conduction was observed at 10 K. Thus, we believe that the phononic crystal concept is not bound to only very low temperatures and manipulation of heat transport using the wave nature of phonons is within reach. Further miniaturization will keep broadening the working temperature range of phononic crystals until the expansion of thermal engineering to wave regime completely changes thermal management, in the way that wave optics revolutionized the manipulations of light.

## Methods

**Sample fabrication.** Each structure is patterned in a region of 20 × 40 µm$^2$. The structures were prepared on commercially available (100) silicon-on-insulator (SOI) wafers. The top silicon layer is nominally doped with boron ($10^{-15}$ cm$^{-3}$) and its thickness is 145 nm. The root mean square roughness has been measured by AFM and was less than 0.5 nm. The thickness of the buried oxide is 1 µm. On the pattern-free SOI substrate, we spin-coated resist (ZEP520-A7, positive), and 4 × 4 µm$^2$ squares were drawn by electron-beam lithography, followed by the development of the resist. A 125 nm-thick aluminum layer was then deposited by electron-beam assisted evaporation before the resist was removed. Immediately afterwards, the resist was spin-coated again. All the structures were patterned by electron-beam lithography aligned with the previously deposited aluminum pads. The resist was used as a mask for the pattern transfer into the silicon. This step was performed by reactive ion etching (Oxford instruments PlasmaLab 100) using a mixture of SF$_6$ and O$_2$ for 20 seconds. After the sample cleaning, the buried oxide layer was removed with diluted hydrofluoric acid vapour in a commercial system, which included a heated plate on which the sample was positioned. This etching method provides good control over the etching speed and allows for a stiction-free buried oxide removal. The size parameters of the holes and beams were validated via scanning electron microscopy (SEM).

**Optical measurement.** The optical measurement system and the method used to obtain the thermal conductivity are detailed in a previous work[24], but were improved to include an automatic exponential fitting of the measurement curve based on the least-squares method. The probe laser is a continuous wave laser diode of 785 nm in wavelength.

Similarly, the pump is a laser diode (642 nm) but pulsed at the frequency of 1 kHz with pulse durations of 1 μs. Both lasers are focused on the sample, passing through the same 40× microscope objective with the aperture of 0.6. The sample itself is mounted in a He-flow cryostat (Oxford Instruments) and measurements are done under vacuum with a pressure below $10^{-2}$ Pa at 300 K and $10^{-4}$ Pa at 4 K to avoid any convection issues. Heat losses through low pressure gas conduction between the structures and the substrate are less than 1 nW, hence negligible. The temperature increase inside the phononic crystal was estimated by comparing the reflectivity change with values of the thermoreflectance coefficient from the literature[28]. This increase is estimated to be less than 2 K and we have experimentally verified that the pump power is in a range in which the decay rate does not vary. The input power is typically within the 200 – 400 nW range. Radiation losses are estimated to be at most 3 nW, which accounts for ~1% of the input power. During measurements at 4 K, the heat load from ambient radiation is approximately 20 nW on a single structure, and does not change our conclusions since it does not depend on disorder. After reflection, the intensity of the probe beam is measured by a silicon photodiode of bandwidth 200 MHz connected to an oscilloscope of bandwidth 1 GHz (Tektronix), which performs an average of the signal over $10^4$ waveforms. The signal is further box-averaged to improve the signal-to-noise ratio and its normalized decay is fitted by an exponential decay curve using the least-squares methods. The dimensions of the structures are measured by SEM and used to create a FEM model (COMSOL MULTIPHYSICS), which virtually reproduces the experiment. The decay times obtained from the simulations for different values of the thermal conductivity are then compared to the experimental decay to extract the measured value of thermal conductivity. The inaccuracy of thermal conductivity measurements is estimated by taking into account the standard deviation on the measurement of the decay rate, the inaccuracies of SEM measurements and FEM analysis, as explained in detail elsewhere[29].

**Acknowledgments** The authors acknowledge the support of the Project for Developing Innovation Systems of the Ministry of Education, Culture, Sports, Science and Technology (MEXT), Japan; of JSPS KAKENHI (grant no. 25709090, 15H05869, and 15K13270); and of JST PRESTO program. The authors thank J. Tatebayashi for his help with SEM observations, and K. Hirakawa, H. Han, and J. Shiomi for fruitful discussions.

**Contributions**

J.M and R.Y. designed and fabricated the samples. J.M performed the measurements. R.A. conducted the analysis and wrote the paper. A.R. performed the MC simulations. S.V. contributed to the analysis and discussion. M.N. supervised the entirety of the work. All authors contributed to the analysis and discussion of the results.

**Author Information** Correspondence and requests for materials should be addressed to M. N. (nomura@iis.u-tokyo.ac.jp) and J.M. (jmaire@iis.u-tokyo.ac.jp).

**Competing financial interests**

The authors declare no competing financial interests.